\documentclass{article}

\usepackage{amsmath,amssymb,amsfonts} 
\usepackage{color}                    

\usepackage{graphicx} 
\usepackage{subfigure} 

\usepackage{natbib}

\usepackage{algorithm}

\usepackage{bm}

\usepackage{hyperref}

\usepackage[accepted]{icml2013}

\icmltitlerunning{Dynamic Covariance Models for Multivariate Financial Time Series}

\begin{document} 

\allowdisplaybreaks

\twocolumn[
\icmltitle{Dynamic Covariance Models for Multivariate Financial Time Series}

\icmlauthor{Yue Wu}{yw289@cam.ac.uk}
\icmlauthor{Jos\'e Miguel Hern\'andez-Lobato}{jmh233@cam.ac.uk}
\icmlauthor{Zoubin Ghahramani}{zoubin@eng.cam.ac.uk}
\icmladdress{University of Cambridge, Department of Engineering,
            Cambridge CB2 1PZ, UK}

\icmlkeywords{boring formatting information, machine learning, ICML}

\vskip 0.3in
]

\begin{abstract} 
The accurate prediction of time-changing covariances is an important problem in
the modeling of multivariate financial data. However,
some of the most popular models suffer from a) overfitting problems and multiple local optima,
b) failure to capture shifts in market conditions and c) large computational costs.
To address these problems we introduce a novel dynamic model for time-changing covariances.
Over-fitting and local optima are avoided by following a Bayesian approach
instead of computing point estimates.
Changes in market conditions are captured by assuming a diffusion process in parameter values, 
and finally computationally efficient and scalable inference is performed using particle filters. 
Experiments with financial data show excellent performance of the proposed method with respect
to current standard models.
\end{abstract} 

\section{Introduction}

Univariate financial returns are heteroscedastic, that is,
the volatility or standard deviation of financial returns is not constant, but changes with time \citep{Cont2001}.
Several univariate models have been proposed to capture this property.
The best known are the Autoregressive Conditional Heteroscedasticity model (ARCH) \citep{engle1982autoregressive}
and its extension, the Generalized Autoregressive Conditional Heteroscedasticity model (GARCH) \citep{bollerslev1986generalized}.
This topic has recently received attention in the machine learning community, with the development of
copula processes \cite{NIPS2010_0784} and heteroskedastic Gaussian processes \cite{Lazaro2011}.

Multivariate financial returns exhibit similar patterns.
Moreover, besides time-dependent volatilities, they also display
time-dependent correlations \citep{Patton2006}. Covariances are the product of
variances and correlations. Therefore these temporal dependencies are likewise present 
in covariance matrices. To capture these properties
\citet{engle1995multivariate} proposed a multivariate extension of GARCH called BEKK,
based on synthesized work of Baba, Engle, Kraft, and Kroner.
An alternative to BEKK are stochastic volatility models, including
recent non-parametric models based on generalized Wishart processes \cite{Wilson11,fox2011bayesian}.
In practice BEKK performs similarly to the generalized Wishart processes for modeling financial data.
However, BEKK is known to suffer from the following disadvantages:

\begin{enumerate}
\item The parameters of BEKK are usually fit by maximum likelihood. 
The large number of parameters in BEKK and local maxima in the likelihood function 
often lead to overfitting.
\item The parameter values in BEKK are constant. However,
financial markets are dynamic and market conditions change with time.
BEKK does not naturally capture these shifts in market conditions. 
\item Finally, maximum likelihood fit of the parameters of BEKK involves solving a non-linear
optimization process which is computationally expensive and infeasible in high-dimensions.
\end{enumerate}

To address the difficulties mentioned above, we present a novel
dynamic model for describing time-dependent covariance matrices which extends BEKK as follows:
Instead of computing point estimates, as in BEKK, we follow a fully Bayesian approach
and compute posterior probabilities for the parameter values.
This reduces the detrimental effect of multiple maxima in the likelihood function and
limits overfitting problems. In addition to this, the new dynamic model incorporates a diffusion process for parameter values.
At each point in time, every parameter is slightly modified by a random perturbation.
These perturbations allow the model to adapt its parameters to changes in market conditions.
Finally, Bayesian inference is performed using a regularized auxiliary particle filter \citep{liu1999combined}.
This technique is very efficient in terms of computational cost and allows our method to scale up to high dimensions.

The performance of the new dynamic model for time-changing covariance matrices is
evaluated in a series of experiments with real financial returns. 
The proposed model is compared with the standard BEKK model and a variant of BEKK that assumes multivariate Student's $t$ innovations.
Finally, we compare the proposed model against the generalized Wishart processes.
Overall, the new dynamic model for time-varying covariance matrices obtains the best predictive performance.

The rest of the document is organized as follows:
Section \ref{sec:Review} describes standard heteroscedastic models such as GARCH and BEKK.
Section \ref{sec:BMDC} introduces our novel Bayesian Multivariate Dynamic Covariance model (BMDC).
Section \ref{sec:relatedwork} reviews current models for dynamic covariances in machine learning.
Section \ref{sec:inferenceMethods} describes the particle filter algorithm for making inference in BMDC.
Experiments comparing BMDC with BEKK are included in Section \ref{sec:results} and
additional experiments comparing BMDC with the generalized Wishart process are included in Section \ref{sec:resultsGWP}.
Finally, Section \ref{sec:summary} contains the conclusions of the study.

\section{Review of GARCH and BEKK}\label{sec:Review}

\allowdisplaybreaks

GARCH is the standard time-series model for univariate heteroscedastic data.
GARCH assumes Gaussian noise or innovations and produces a sequence of time-varying variances
$\sigma_{t}^{2}$ that follow an Autoregressive and Moving Average (ARMA) process
with autoregression on $p$ previous variance values and moving average on $q$ previous squared time-series values:
\begin{align}
	x_{t} &\sim \mathcal{N}(0,\sigma_{t}^{2})\,,\\ 
	\sigma_{t}^{2} &= \alpha_{0} + \sum_{j=1}^{q} \alpha_{j} x_{t-j}^{2} + \sum_{i=1}^{p} \beta_{i} \sigma_{t-i}^{2}\,. \label{eq:garch}
\end{align} 
The generative model is flexible and can produce a variety of clustering behavior of high and low volatility periods for
different settings of the model coefficients, $\alpha_1,\ldots,\alpha_p$ and $\beta_1,\ldots,\beta_q$.
Maximum likelihood is used to learn these coefficients with
$p$ and $q$ usually set to 1 to reduce overfitting problems.

BEKK \citep{engle1995multivariate} is a popular multivariate extension of GARCH, where the dynamic covariance matrix for the data follows an ARMA process:
\begin{align}
	\mathbf{x}_{t} &\sim  \mathcal{N}(0, \bm \Sigma_{t})\,,\label{test1}\\
	\bm \Sigma_{t} &=  \mathbf{C}^{\top} \mathbf{C} +
	\sum_{j=1}^{q} \mathbf{B}_{j}^{\top} \mathbf{x}_{t-j} \mathbf{x}_{t-j}^{\top} \mathbf{B}_{j} +
	\sum_{i=1}^{p} \mathbf{A}_{i}^{\top} \bm \Sigma_{t-i} \mathbf{A}_{i}\label{test2}
\end{align}
where $\mathbf{A}_{i}$ and $\mathbf{B}_{j}$ are $d \times d$ coefficient matrices for $d$ dimensional data and
$C$ is a triangular matrix with $d(d+1)/2$ non-zero entries. Therefore BEKK($p$,$q$) is a highly parameterized
model with a total of $(p+q) d^2 + d(d+1)/2$ parameters.

Restricted versions of BEKK are used in practice to mitigate overfitting problems.
The order parameters $p$ and $q$ are set to 1 and
the matrices $\mathbf{A}_1$ and $\mathbf{B}_1$ are constrained to be diagonal.
The expression for $\bm \Sigma_{t}$ is now
\begin{align}
	\bm \Sigma_{t} &=  \mathbf{C}^{\top} \mathbf{C} + \mathbf{B}  \mathbf{x}_{t-1} \mathbf{x}_{t-1}^{\top} \mathbf{B} +
	\mathbf{A} \bm \Sigma_{t-1} \mathbf{A}\,.  \label{eq:diagBEKK}
\end{align}
This diagonal BEKK model \citep{engle1995multivariate} has only $2d+ d(d+1)/2$ parameters. We will use this model
as the baseline for comparison because it often has better predictive performance than other versions of BEKK
and its computational cost is also lower. Subsequent references to BEKK will mean the diagonal BEKK model shown in (\ref{eq:diagBEKK}).

Standard BEKK has Gaussian innovations. However, there is strong evidence that financial returns are heavy-tailed.
\citet{harvey1992unobserved} and \citet{fiorentini2003maximum} incorporate heavy-tails in BEKK by modeling
$\mathbf{x}_{t}$ using a multivariate Student's $t$ distribution with $\nu$ degrees of freedom, zero mean, and scale matrix $\mathbf{S}_{t}$:
\begin{align}
	\mathbf{x}_{t} &\sim  \mathcal{S}(\nu, 0, \mathbf{S}_{t})\,, \\
	p(\mathbf{x}_{t}) & = \frac{\Gamma[(\nu+d)/2] [1+\frac{1}{\nu} \mathbf{x}_{t}^{\top}\mathbf{S}_{t}^{-1}
	\mathbf{x}_{t}]^{-(\nu+d)/2}}{\Gamma(\nu/2) (\nu\pi)^{d/2} \abs{\mathbf{S}_{t}}^{1/2}}\,, \\
	\mathbf{S}_{t} &= \frac{\nu-2}{\nu}\bm \Sigma_{t}\,.
\end{align}
The degrees of freedom, $\nu$, should be larger than two for $\mathbf{x}_t$ to have a well defined covariance.
Finally, the expression for $\mathbf{S}_{t}$ ensures the expected covariance of
$\mathbf{x}_{t}$ is $\bm \Sigma_{t}$.  We will refer to the diagonal BEKK model with Student's $t$ innovations as BEKK-T.

\begin{figure*}
    \centering
    \includegraphics[scale=.365, trim=180mm 160mm 50mm 10mm, clip ]{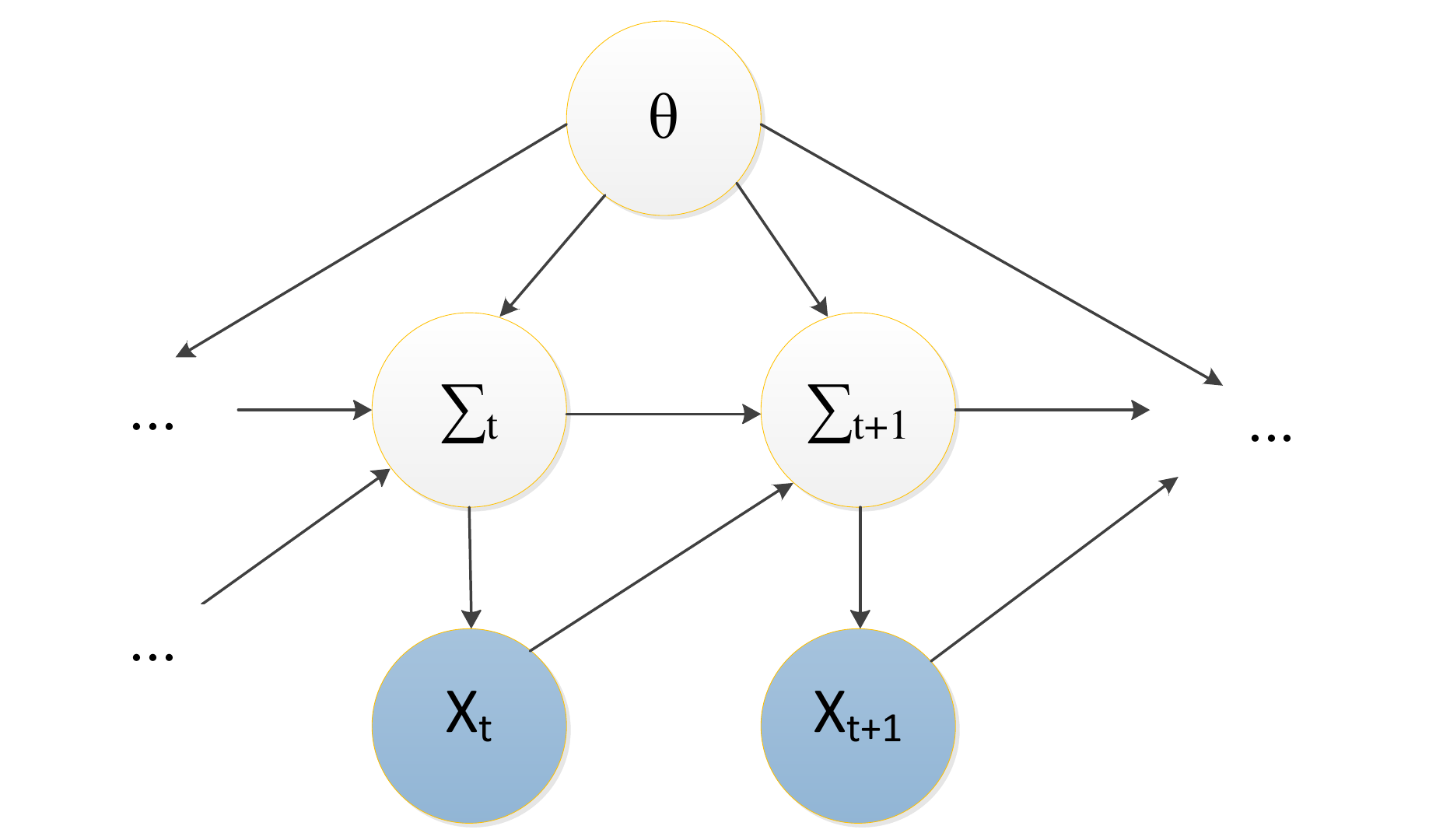}\hspace{0.25cm}
    \includegraphics[scale=.365, trim=0mm 140mm 0mm 10mm, clip ]{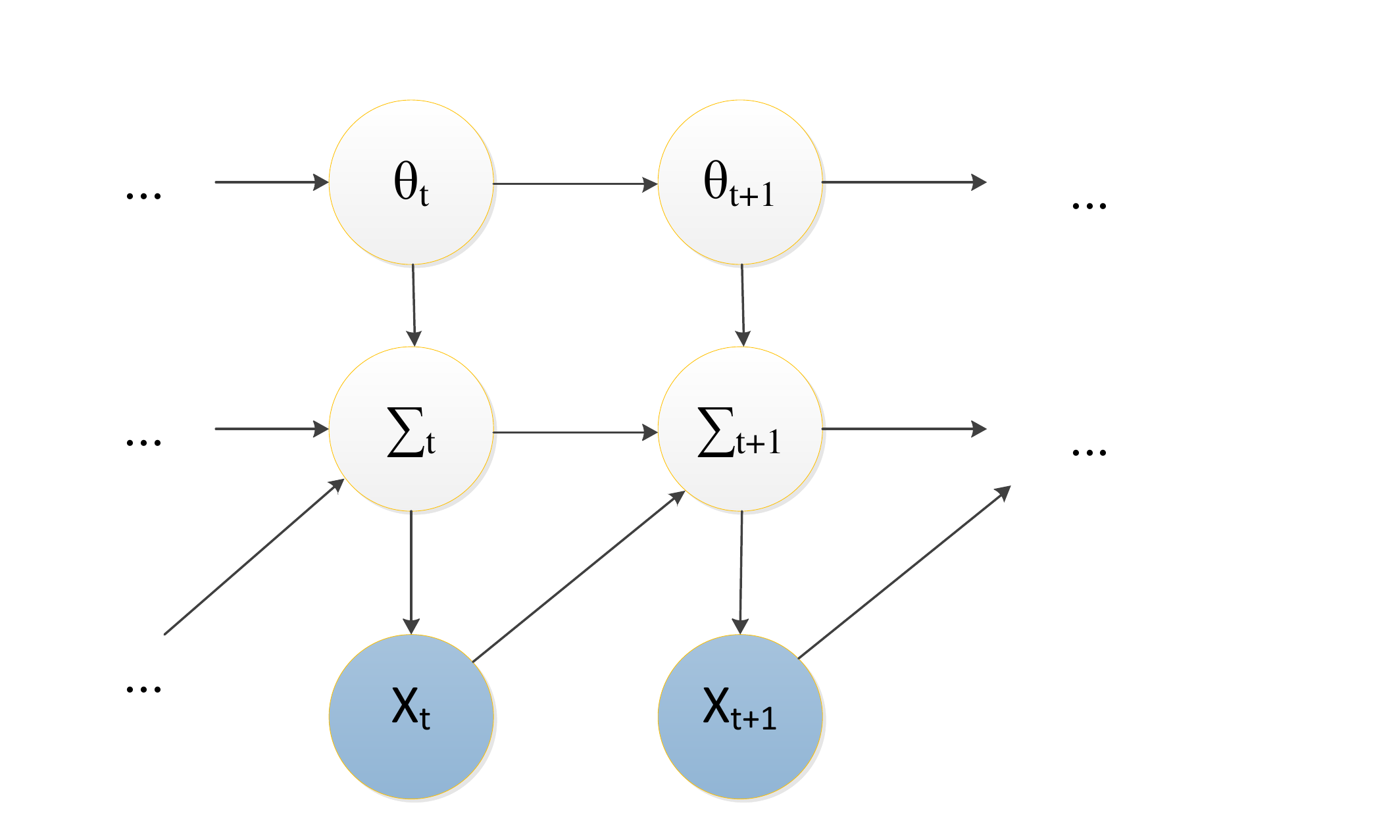} 
    \caption{Left, graphical model for BEKK. The parameters $\theta=(\mathbf{A},\mathbf{B},\mathbf{C})$ are in this case static.
    Right, graphical model for BMDC with time-varying parameters $\mathbf{\theta}_{t}=(\mathbf{A}_{t}, \mathbf{B}_{t}, \mathbf{C}_{t})$. BMDC naturally
    captures changes in market conditions.}
    \label{fig:bayesian}
\end{figure*}

\section{Bayesian Multivariate Dynamic Covariance Models} \label{sec:BMDC}

A major limitation of BEKK is that the parameter matrices $\mathbf{A}$, $\mathbf{B}$ and $\mathbf{C}$ are assumed to be constant.
This is unrealistic for financial data, where market fundamentals are expected to change with time.
A heuristic solution is to run BEKK over different windows of historical data. 
The problem is then how to choose the window size, with trade-offs between 
noisy but reactive estimates of the model parameters for small windows and
more stable but constant parameter estimates for large windows.

As a more principled approach to capture changes in market dynamics,
we introduce a novel Bayesian Multivariate Dynamic Covariance model (BMDC) with time-varying parameters.
For this, we replace (\ref{eq:diagBEKK}) by
\begin{align}
	\bm \Sigma_{t} &= \mathbf{C}_{t}^{\top} \mathbf{C}_{t} + \mathbf{B}_{t}
	\mathbf{x}_{t-1} \mathbf{x}_{t-1}^{\top} \mathbf{B}_{t} + \mathbf{A}_{t} \Sigma_{t-1} \mathbf{A}_{t}\,. \label{eq:bayesDynamic} 
\end{align}
where $\mathbf{C}_t$, $\mathbf{B}_t$ and $\mathbf{A}_t$ are time-dependent matrices.
Let $\mathbf{\theta}=\{\mathbf{A},\mathbf{B},\mathbf{C}\}$ and $\mathbf{\theta}_t=\{\mathbf{A}_t,\mathbf{B}_t,\mathbf{C}_t\}$.
Figure \ref{fig:bayesian} shows the corresponding graphical models for BEKK and BMDC.

In BMDC, the dynamic parameters in $\theta_t$ follow a diffusion process
in which $\theta_{t+1}$ is obtained by adding a small random perturbation to the parameters in $\theta_t$.
Since $\mathbf{A}_{t}$ and $\mathbf{B}_{t}$ are diagonal, let $\mathbf{a}_{t}$ and $\mathbf{b}_{t}$ denote $d$-dimensional vectors with the
diagonal elements of these matrices and let $\mathbf{c}_{t}$ be the vector with the upper triangular terms of $\mathbf{C}_{t}$.
Then we specify the following diffusion process for $\mathbf{a}_{t}$, $\mathbf{b}_{t}$ and $\mathbf{c}_{t}$:
\begin{align}
	\mathbf{a}_{t} &\sim \mathcal{N}(\mathbf{a}_{t-1}, \alpha^2 I)\,,\label{eq:bayesConditionalPrior1}\\
	\mathbf{b}_{t} &\sim \mathcal{N}(\mathbf{b}_{t-1}, \beta^2 I)\,,\label{eq:bayesConditionalPrior2}\\
	\mathbf{c}_{t} &\sim \mathcal{N}(\mathbf{c}_{t-1}, \gamma^2 I)\,,\label{eq:bayesConditionalPrior3}\\
	\alpha &\sim \mathcal{N}(\kappa, \tau)\,, \quad
	\beta \sim \mathcal{N}(\kappa, \tau)\,, \quad
	\gamma \sim \mathcal{N}(\kappa, \tau)\,. \label{eq:bayesDynamicJitter}
\end{align}
where the hyper-parameters $\alpha$, $\beta$ and $\gamma$ control the amount of drift in the system. 
In practice, vague priors are chosen for the value of these hyper-parameter,
with mean prior drift set to zero, $\kappa=0$. 
$\tau$ was set to $0.005^2$ so that $A$, $B$ and $C$ can move up to $0.005\sqrt{N}$, where $N$ is the number of observations. 
Other values of $\tau$ were tried with little difference in predictive performance.
The prior for the initial state $\mathbf{a}_{0}$, $\mathbf{b}_{0}$ and $\mathbf{c}_{0}$ is also vague, taking into account the
constraint $(|\mathbf{A}_{0}\mathbf{A}_{0}|+|\mathbf{B}_{0}\mathbf{B}_{0}|) \leq 1$ so that $\bm \Sigma_t$ does not diverge,
where $|\cdot|$ calculates the determinant of a matrix.

An important property of BMDC is that its predictive distribution is heavy-tailed.
This is desirable as financial time series have fat tails \cite{Cont2001}.
The heavy-tails in the predictions of BMDC arise because
\begin{align}
p(\mathbf{x}_{t}^{\star} | \bm \Sigma_{t-1}, \mathbf{x}_{t-1}) = 
 \int p(\mathbf{x}_{t}^{\star} | \bm \Sigma_{t})p(\bm \Sigma_{t}| \bm \Sigma_{t-1},\mathbf{x}_{t-1}) d \bm \Sigma_{t} \,.\label{eq:posteriorPredictive}
\end{align}
where the distribution of $\bm \Sigma_{t} | \bm \Sigma_{t-1}, \mathbf{x}_{t-1}$
is obtained by integrating \eqref{eq:bayesDynamic} with respect to the posterior for
$\theta_{t}=(\mathbf{A}_{t}, \mathbf{B}_{t}, \mathbf{C}_{t})$.
Since $p(\mathbf{x}_{t}^{\star} | \bm \Sigma_{t})$ is Gaussian, (\ref{eq:posteriorPredictive})
is an infinite mixture of multivariate Gaussians and it will generally be heavy-tailed.
This implies that the predictive density in the BMDC model will also have heavy tails.

In addition to BMDC, we propose a variant of this model with
Student's $t$ innovations, which we denote BMDC-T.
In this case, a vague log-normal prior is placed on the parameter $\nu$ corresponding to the number of degrees of freedom of the Student's $t$ innovations:
\begin{align}
\log (\nu - 2) \sim \mathcal{N}(0,\sigma_{\nu}^{2})\,. \notag
\end{align}

\section{Related Work}\label{sec:relatedwork}

The GARCH and BEKK models described in Section \ref{sec:Review} constitute the most popular family of volatility models.
They are characterized by the latent covariance matrix $\bm \Sigma_t$ being dependent on both its most recent past value $\bm \Sigma_{t-1}$
and the previous time-series observation $\mathbf{x}_{t-1}$.
Another class of models is the family of Stochastic Volatility models \citep{harvey1994multivariate,chib2009multivariate,gourieroux2009wishart,philipov2006factor}.
In this case, $\bm \Sigma_t$ is conditionally independent of $\mathbf{x}_{t-1}$ given its previous value $\bm \Sigma_{t-1}$.
That is, graphically, in a stochastic volatility model there will be no arrow from node $\mathbf{x}_t$ to node $\bm \Sigma_{t+1}$ in the
graphs shown in Figure \ref{fig:bayesian}. However, $\bm \Sigma_{t}$ will still be conditionally independent 
of $\bm \Sigma_{k}$ for $k < t-1$ and $k > t+1$ given $\bm \Sigma_{t+1}$ and $\bm \Sigma_{t-1}$. 
This latter condition does not hold in recent generalizations of these models based on generalized Wishart processes (GWP)
\citep{fox2011bayesian,Wilson11}. In this case, there are dependencies between all the latent covariance matrices
and the model can produce complex non-linear patterns in the evolution of covariance matrices.
However, it is not clear that such flexibility is necessary for successfully modeling financial data.
Our experiments show that BMDC outperforms GWP on this task, which confirms that this is not the case.

\section{Inference with Particle Filters} \label{sec:inferenceMethods}

Inference for the proposed Bayesian models is performed using particle filters \citep{doucet2001sequential}. 
Particle filters seem a natural choice to do online inference for these non-linear and non-Gaussian sequential models. 
BMDC has Gaussian likelihoods, but is non-linear in the model parameters $\mathbf{A}_t$, $\mathbf{B}_t$ and $\mathbf{C}_t$.
Furthermore, particle filters can easily accommodate models with Student's $t$ innovations, such as BMDC-T. 

We analyzed several particle filtering methods, including 
Resample-Move \citep{gilks2001following}, 
Regularized Sequential Importance Sampling \citep{musso2001improving}, 
Regularized Sequential Importance Resampling \citep{gordon1993novel}, and 
Regularized Auxiliary Particle Filter (RAPF) \citep{liu1999combined} to perform inference in the proposed models. 
RAPF, a hybrid version of regularized \citep{musso2001improving} and auxiliary
particle filters \citep{pitt1999filtering}, exhibited the best performance in terms of predictiveness.
This agrees with results given by \citet{casarin2009online}. 
An implementation of the RAPF for BMDC is shown in Algorithm 1.
A detailed description of the algorithm is given in the following paragraph.

\citet{liu1999combined} introduced the RAPF,
which combines the Regularized Particle Filter with the Auxiliary Particle Filter. 
The Regularized Particle Filter allows joint inference of the diffusion
hyper-parameters, $\alpha$, $\beta$, $\gamma$, and the hidden states, $\mathbf{A}_{t}$, $\mathbf{B}_{t}$ and $\mathbf{C}_{t}$.
This method explores the posterior distribution of the model parameters by taking kernelized steps sequentially, thereby avoiding particle death. 
The shrinkage kernel, (17) and (19), avoids over-dispersion problems in standard
Regularized Particle Filters and improves prediction accuracy. 
The shrinkage kernel retains the previous mean, while the posterior variance does not diverge:
\begin{align}	
	\mathbb{E}(\mathbf{\theta}_{t}) &= \bar{\mathbf{\theta}}_{t-1} \label{eq:shrinkageMeanConsistency}\\
	\Cov(\mathbf{\theta}_{t}) &= a^{2}V_{t} + (1-a^{2}) V_{t} = V_{t} \label{eq:shrinkageCovConsistency}
\end{align}

\includegraphics[scale=1]{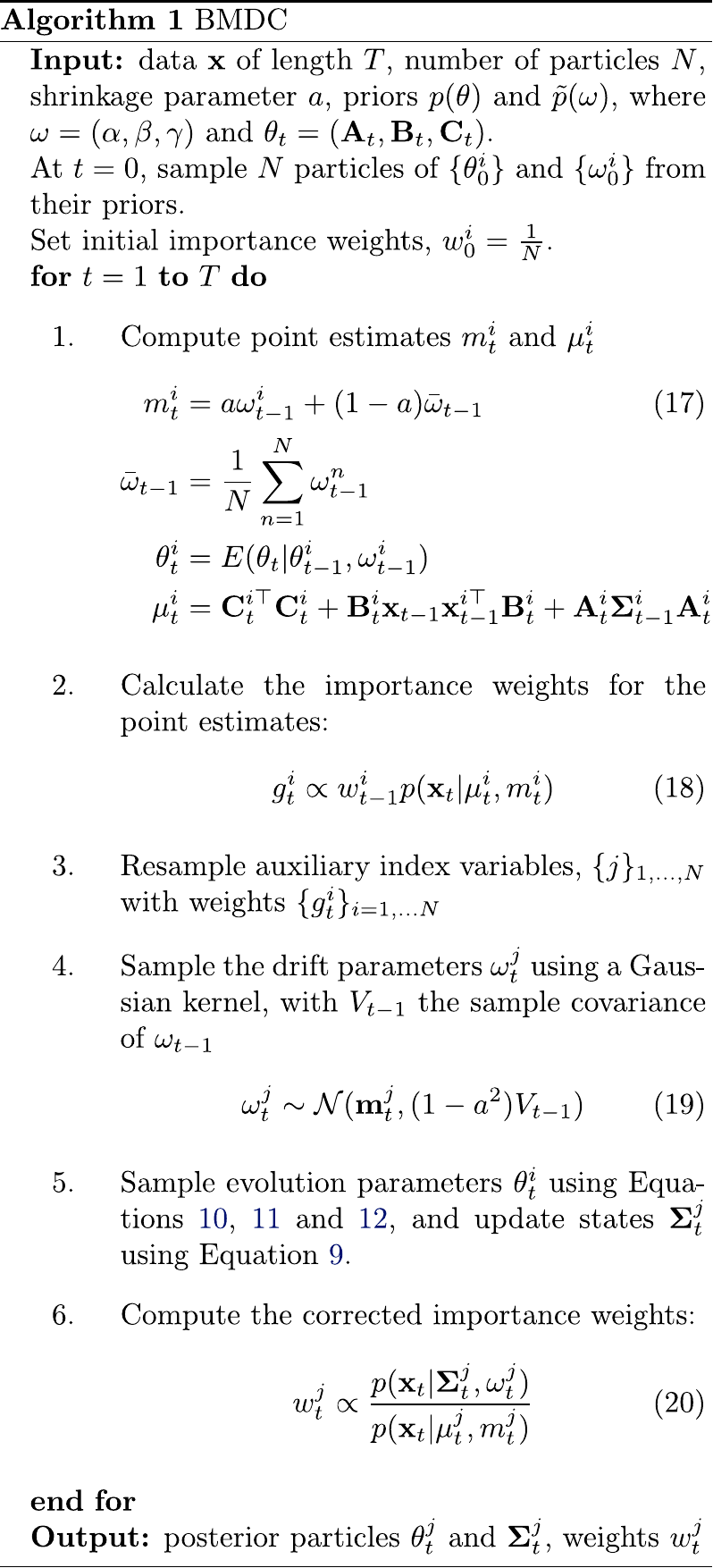}

The Auxiliary Particle Filter part of the algorithm can be viewed as interchanging the
importance sampling and resampling steps in traditional particle filters such as Sequential Importance Resampling. 
The resampling, (18), is performed on $\mu_{t}^{i}$,
which are predicted estimates of $\bm \Sigma_{t}$ given particles from the previous time step.
Importance samples are then generated from the resampled point estimates. 
Particle diversity is maintained if the predicted estimates from the previous step are close to the true state. 
Auxiliary particle filtering is less sensitive to outliers and works well when the predicted estimates are good representations of the unknown states.


There are two algorithm parameters in Algorithm 1.
The number of particles $N$ and the shrinkage parameter $a$.
We followed \citet{liu1999combined} and fixed $a=0.95$.
This leads to a little shrinkage and avoids parameter variance exploding.
If shrinkage is large ($a=0$) then the particle filter would propose 
from a less heavy-tailed proposal distribution,
with the parameters set to be their empirical means from the previous step. 
When the empirical means are inaccurate, then the proposed particles will not be representative of the hidden state.
The number of particles, $N$, can affect the accuracy of the predictions, and is investigated in Section \ref{sec:results}.
The computational complexity of the algorithm is $O(N d^{3})$ at each time step, where $d$ is the dimension of the data,
since importance sampling at each step requires $N$ likelihood computations, each one with cost $O(d^{3})$.

\section{Experiments for BMDC vs. BEKK} \label{sec:results}

We evaluated the performance of BMDC, BEKK and their Student's $t$ variants in several experiments with multivariate financial time series. 
The high computational cost of BEKK limited the maximum number of different financial assets in each of the analyzed series to five.
We considered daily foreign exchange (FX) and daily equity returns, as well as intraday FX returns.
The daily FX time series contain a total of 780 returns from January 2008 to January 2011. 
The daily equity time series contain 3000 returns from Jan 2000 to Dec 2011.
The intraday FX time series consist of 5000 five-minute returns, which covered approximately the first six trading months of 2008.
The price data were pre-processed to eliminate spurious prices. 
In particular, we eliminated prices corresponding to times when markets were closed or not liquid.
All the time series were standardized to have zero mean and unit standard deviation.

To illustrate the usefulness of considering time-varying parameter values,
we compared the predictive performance of BMDC and BMDC-T with 
BEKK and BEKK-T.
In addition, we compared the execution times of i) the RAPF method used by the Bayesian models and ii) 
the maximum likelihood estimation method used by the BEKK models.
Finally, we studied the sensitivity of RAPF to the number of particles used.

The performance of each method is measured in terms of the predictive
log-likelihood on the first return out of the training set.
During the experiments, each method receives an initial time-series of length 50.
The different methods are trained on that data and then a one-step forward prediction is made.
The predictive log-likelihood is evaluated on the next observation out of the training set.
Then the training set is augmented with the new observation and the
training and prediction steps are repeated again.
The process is repeated sequentially until no further data is received.

\newcommand{\ica}{\hspace{0.1cm}}
\renewcommand{\arraystretch}{0.85}

\begin{table}
\centering
\caption{Avg. Predictive Likelihood on Daily FX Data} \label{tbl:BMDCDailyFX}
\resizebox{0.475 \textwidth }{!}{
\begin{tabular}{@{\ica}l@{\ica}c@{\ica}c@{\ica}c@{\ica}c@{\ica}c@{\ica}}
Dataset &  BEKK   &  BEKK-T   &  BMDC   &  BMDC-T   \\ 
\hline
EUR & $-1.47$ & $-1.40$ & $-1.36$ & $\mathbf{-1.36}$ \\
AUD & $-1.32$ & $-1.29$ & $-1.25$ & $\mathbf{-1.25}$ \\
BRL & $-1.18$ & $-1.16$ & $-1.15$ & $\mathbf{-1.14}$ \\
GBP & $-1.33$ & $-1.33$ & $\mathbf{-1.31}$ & $\mathbf{-1.31}$ \\
CHF & $-1.44$ & $-1.45$ & $-1.37$ & $\mathbf{-1.37}$ \\
AUD,JPY & $-2.62$ & $\mathbf{-2.56}$ & $-2.57$ & $\mathbf{-2.56}$ \\
EUR,CHF & $-2.15$ & $\mathbf{-2.04}$ & $-2.07$ & $-2.05$ \\
EUR,GBP & $-2.47$ & $-2.44$ & $\mathbf{-2.41}$ & $\mathbf{-2.41}$ \\
BRL,EUR & $-2.52$ & $-2.46$ & $\mathbf{-2.43}$ & $\mathbf{-2.42}$ \\
ZAR,AUD & $-2.34$ & $-2.27$ & $\mathbf{-2.27}$ & $\mathbf{-2.26}$ \\
CHF,EUR,JPY & $-3.75$ & $-3.56$ & $-3.27$ & $\mathbf{-3.20}$ \\
JPY,AUD,NZD & $-6.82$ & $-3.77$ & $-3.24$ & $\mathbf{-3.20}$ \\
BRL,AUD,ZAR & $-3.38$ & $-3.31$ & $-3.25$ & $\mathbf{-3.23}$ \\
TWD,JPY,KRW & $-7.43$ & $-5.71$ & $-3.61$ & $\mathbf{-3.57}$ \\
CAD,MXN,BRL & $-3.43$ & $-3.33$ & $-3.19$ & $\mathbf{-3.17}$ \\
JPY,AUD,EUR,CHF & $-4.71$ & $-4.75$ & $-4.21$ & $\mathbf{-4.11}$ \\
BRL,MXN,CAD,AUD & $-4.67$ & $-4.30$ & $-4.09$ & $\mathbf{-4.01}$ \\
BRL,ZAR,AUD,NOK & $-4.62$ & $-4.46$ & $-4.22$ & $\mathbf{-4.19}$ \\
JPY,AUD,GBP,EUR,CHF & $-8.94$ & $-6.57$ & $-5.31$ & $\mathbf{-5.16}$ \\
BRL,MXN,CAD,AUD,ZAR & $-5.51$ & $-5.41$ & $-5.14$ & $\mathbf{-5.01}$ \\
NZD,AUD,JPY,EUR,SEK & $-6.33$ & $-5.58$ & $-4.90$ & $\mathbf{-4.80}$ \\
\hline
\end{tabular}
}
\end{table}

\begin{table}
\centering
\caption{Avg. Predictive Likelihood on Intraday FX Data}\label{tbl:BMDCIntradayFX}
\resizebox{0.475 \textwidth }{!}{
\begin{tabular}{@{\ica}l@{\ica}c@{\ica}c@{\ica}c@{\ica}c@{\ica}c@{\ica}}
Dataset &  BEKK   &  BEKK-T   &  BMDC   &  BMDC-T   \\ 
\hline
AUDJPY & $-1.16$ & $\mathbf{-1.09}$ & $\mathbf{-1.09}$ & $\mathbf{-1.09}$ \\
AUDUSD & $-1.19$ & $\mathbf{-1.14}$ & $\mathbf{-1.15}$ & $\mathbf{-1.15}$ \\
EURAUD & $-1.15$ & $\mathbf{-1.10}$ & $\mathbf{-1.10}$ & $\mathbf{-1.10}$ \\
EURCHF & $-1.25$ & $\mathbf{-1.20}$ & $-1.22$ & $-1.22$ \\
EURCZK & $-1.26$ & $\mathbf{-1.11}$ & $-1.16$ & $-1.14$ \\
EURGBP,EURUSD & $-2.57$ & $\mathbf{-2.49}$ & $-2.51$ & $\mathbf{-2.50}$ \\
EURHUF,GBPJPY & $-2.33$ & $-2.20$ & $-2.22$ & $\mathbf{-2.17}$ \\
EURJPY,GBPUSD & $-2.37$ & $\mathbf{-2.27}$ & $-2.29$ & $\mathbf{-2.27}$ \\
EURNOK,NZDUSD & $-2.88$ & $\mathbf{-2.18}$ & $-2.25$ & $-2.20$ \\
EURSEK,USDCAD & $-2.45$ & $\mathbf{-2.32}$ & $-2.35$ & $\mathbf{-2.32}$ \\
EURUSD,USDCHF,USDJPY & $-3.57$ & $-3.28$ & $-3.26$ & $\mathbf{-3.18}$ \\
GBPJPY,USDJPY,USDMXN & $-3.17$ & $-2.56$ & $-2.52$ & $\mathbf{-2.41}$ \\
GBPUSD,USDMXN,USDNOK & $-3.14$ & $-2.84$ & $-2.79$ & $\mathbf{-2.65}$ \\
NZDUSD,USDNOK,USDSEK & $-3.63$ & $-3.41$ & $-3.15$ & $\mathbf{-3.07}$ \\
USDCAD,USDSEK,USDSGD & $-3.78$ & $-3.67$ & $-3.69$ & $\mathbf{-3.61}$ \\
\hline
\end{tabular}}
\end{table}

\subsection{Results for BMDC vs. BEKK}

Table \ref{tbl:BMDCDailyFX} shows the average predictive log-likelihood for BEKK, BEKK-T, BMDC and  BMDC-T
on twenty-one daily FX time series sorted by the dimensionality of the data.
The method with the best predictive performance on each time series is highlighted in bold.
Corresponding results are shown in tables \ref{tbl:BMDCIntradayFX} and \ref{tbl:BMDCDailyEquity}
for fifteen intraday FX and thirty daily equity time series respectively. Overall,
the best performing method is BMDC-T followed by BEKK-T and BMDC. The worst performing method is BEKK.

\begin{figure*}
\centering \includegraphics[scale=.45]{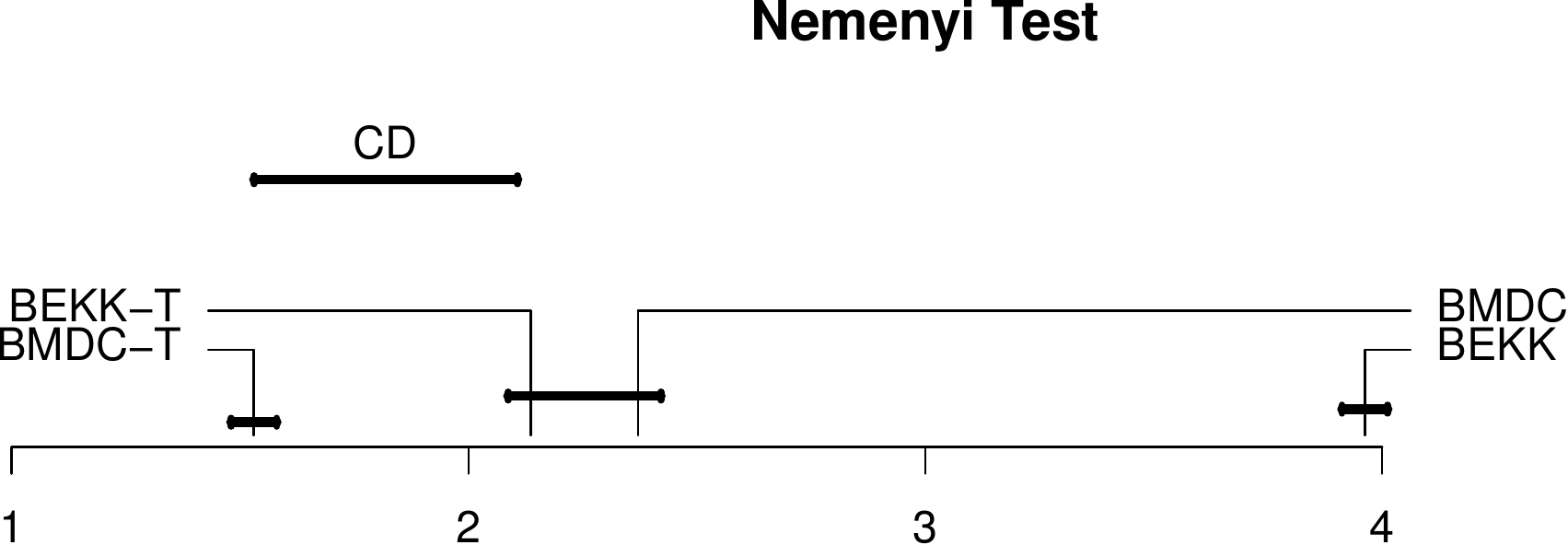} 
\caption{All to all comparison between BMDC-T, BMDC, BEKK-T and BEKK via a Nemenyi test.
The horizontal axis indicates the average rank of each method on the 66 analyzed time series. If the differences in average ranks are
larger than the critical distance (length of the segment labeled CD)
then differences in performance are statistically significant
at $\alpha = 0.05$. In this case, the differences in rank between BMDC-T and all the other
methods are significant.}
\label{fig:bonferroniNew}
\end{figure*}

We perform a statistical test to determine whether differences among BEKK, BEKK-T, BMDC and BMDC-T
are significant. These methods are compared to each other using the multiple comparison approach described by \citet{demšar2006statistical}.
In this comparison framework, all the methods are ranked according to their performance on different tasks.
Statistical tests are then applied to determine whether the differences
among the average ranks of the methods are significant.
In our case, each of the $66=21+15+30$ datasets analyzed represents a different task.
A Friedman rank sum test rejects the hypothesis that all methods
have equivalent performance at $\alpha = 0.05$. Pairwise comparisons between
all the  methods with a Nemenyi test at a 95\% confidence level are summarized in Figure \ref{fig:bonferroniNew}.
The methods whose average ranks across datasets differ more than a critical distance (segment labeled CD in the figure)
show significant differences in performance at this confidence level.
The Nemenyi test confirms that BMDC-T is superior to all the other methods.
Additionally, the dynamic methods are superior to their static counterparts with BMDC outperforming BEKK
and BMDC-T beating BEKK-T. Finally, BMDC is not statistically different from BEKK-T.

\begin{figure}[h]
\centering
\includegraphics[scale=.32]{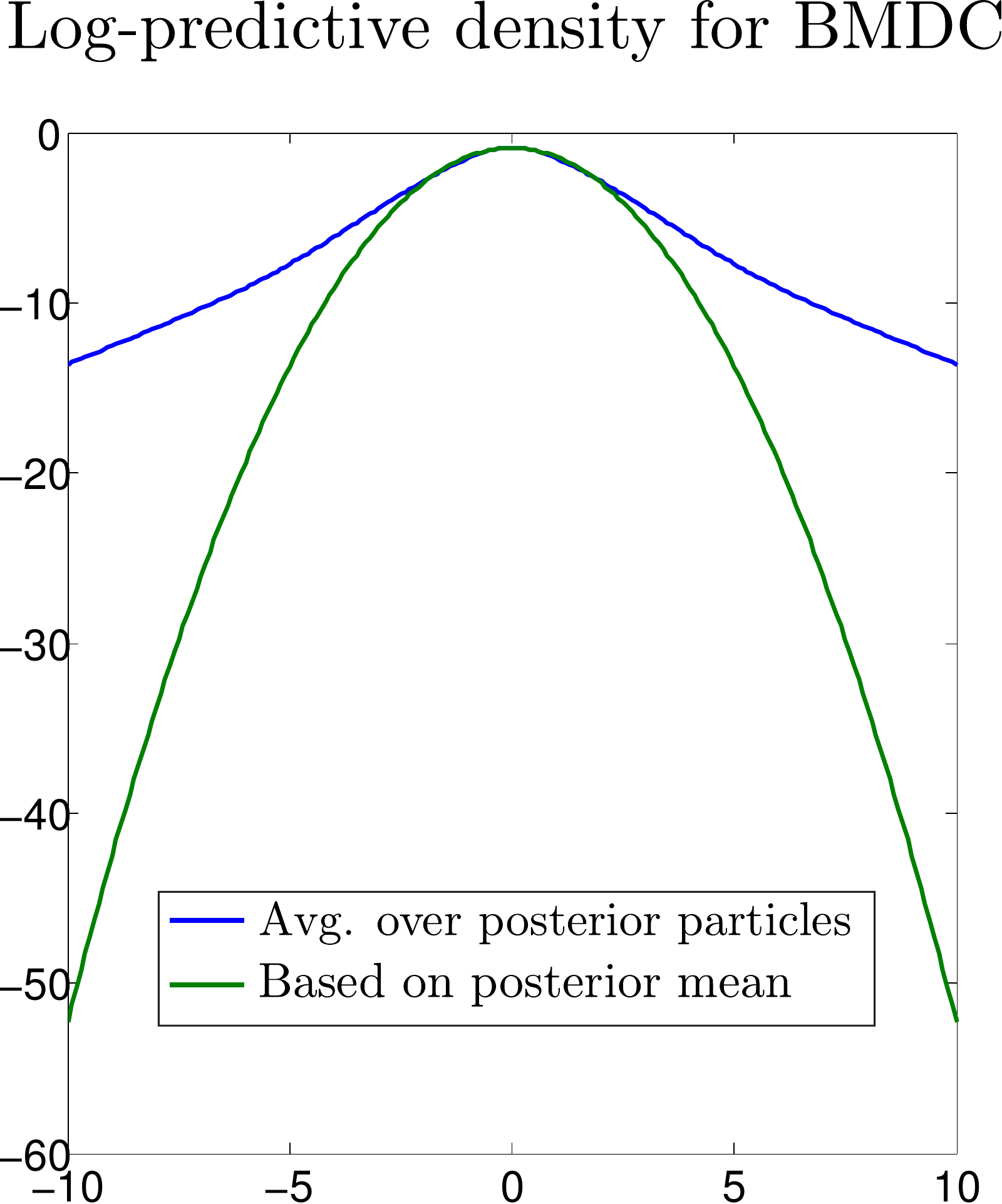} 
\caption{
The log predictive density using the full posterior is
much flatter and thereby heavy-tailed than the one using only the posterior mean for BMDC}\label{fig:dynamicHeavyTailedness}
\end{figure}

The plot in Figure \ref{fig:bonferroniNew} shows that the heavy-tailed models BMDC-T and BEKK-T perform better than the non-heavy-tailed BEKK.
Although BMDC assumes a Gaussian likelihood its posterior predictive distribution is heavy-tailed, as discussed in Section \ref{sec:BMDC}.
Figure \ref{fig:dynamicHeavyTailedness} confirms this by showing the logarithm  of the posterior predictive density
produced by BMDC on a particular instance of the analyzed
time series. To produce this plot, we evaluated $p(\mathbf{x}_t|\theta_t,\bm \Sigma_{t-1},\mathbf{x}_{t-1})$
on a grid of values for one dimension of $\mathbf{x}_{t}$, averaging over the available particles
which approximate the posterior distribution of $\theta_t$ and $\bm \Sigma_{t-1}$.
This is compared to the plot obtained by evaluating $p(\mathbf{x}_t|\theta_t,\bm \Sigma_{t-1},\mathbf{x}_{t-1})$ on 
the posterior mean estimate of $\theta_t$ and $\bm \Sigma_{t-1}$, which is approximated by the empirical mean computed across all the particles.

\begin{figure}[h]
\centering \includegraphics[scale=.36]{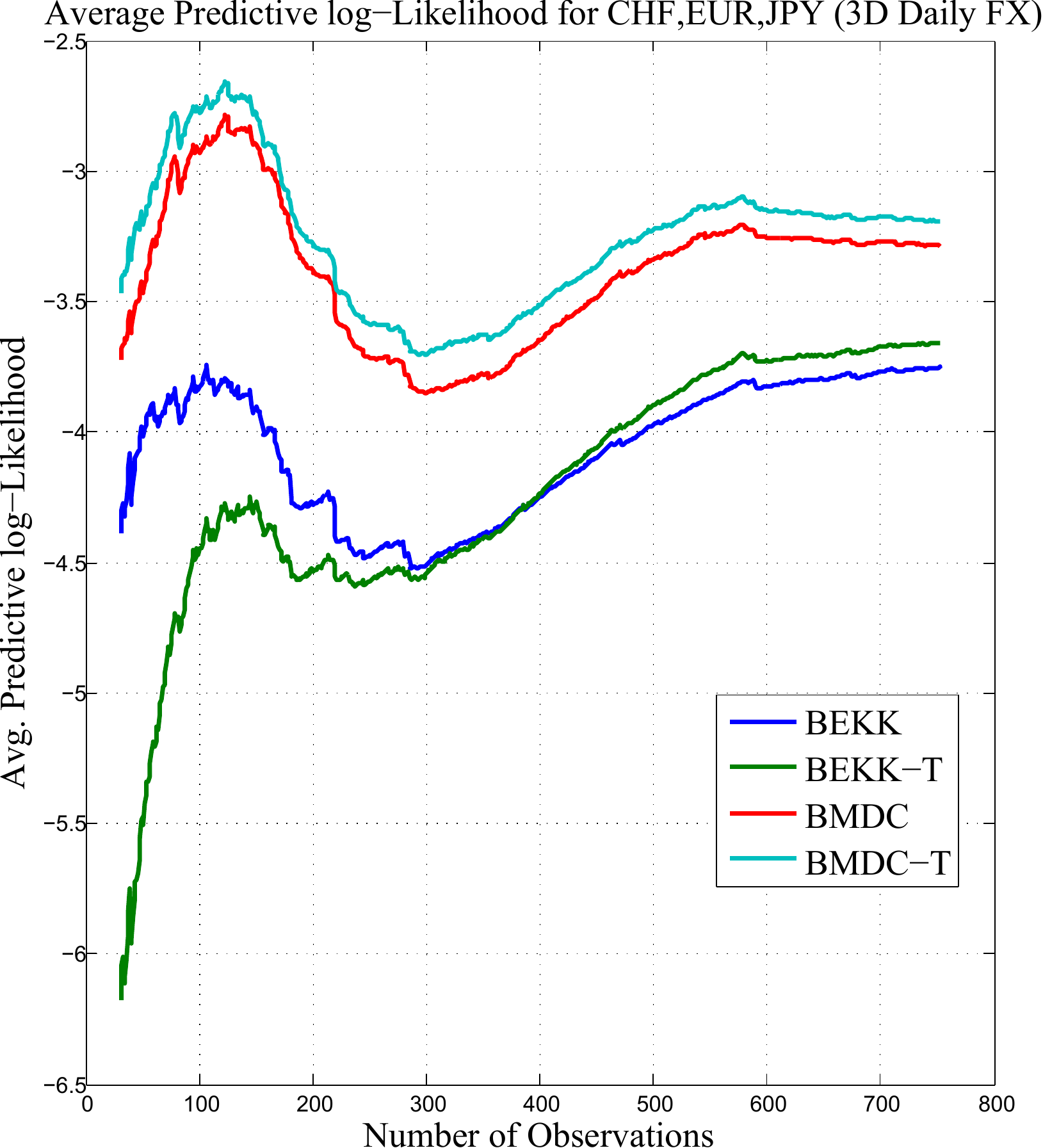}
\caption{Avg. predictive log-Likelihood for each method over number of observations.}
\label{fig:avgPredLik}
\end{figure}

To understand the superior performance of BMDC relative to BEKK,
we plot the average predictive log-likelihood of each method against the number of observations in Figure \ref{fig:avgPredLik}.
The plot shows typical average predictive log-likelihood for a $3D$ Daily FX time series.
BMDC-T is clearly the most predictive method for any number of observations.
BEKK and BEKK-T underperform BMDC and BMDC-T early on, when relatively few observations are available to fit the models.
These two methods are susceptible to overfitting at early stages (especially BEKK-T in this case).
With more data, BEKK-T is less susceptible to overfitting and outperforms BEKK.
However, BEKK and BEKK-T still perform worse than BMDC and BMDC-T when the amount of training data increases.
This confirms that BMDC and BMDC-T are better models for dynamic covariances in financial data,
not only because BEKK and BEKK-T suffer from initial overfitting problems.
As a note of financial interest, the average predictive log-likelihood dips after 150 observations.
This corresponds to the highly turbulent period near the end of 2008 with large swings in financial asset prices resulting in lower predictive log-likelihood.

\begin{table*}
\caption{Sensitivity in the average predictive log-likelihood of BMDC-T to the number of Particles used on Daily FX data.
Execution times in minutes are shown in parentheses. Results of BEKK-T are shown for comparison.}
\label{tbl:timeComparison}
\begin{center}
\resizebox{0.8 \textwidth }{!}{
\begin{tabular}{ p{2cm} p{2cm}p{2cm}p{2cm}p{2cm}p{2.5cm}}
Dataset &  BEKK-T   &  N=1000   &  N=4000   &  N=9000   &  N=25000   \\ 
\hline
1D & $-1.33$  $(  71)$ & $-1.32$  $(   8)$ & $-1.32$  $(  20)$ & $-1.32$  $(  52)$ & $\mathbf{-1.32}$  $( 173)$ \\
2D & $-2.56$  $( 339)$ & $-2.58$  $(   8)$ & $-2.56$  $(  25)$ & $\mathbf{-2.55}$  $(  55)$ & $-2.56$  $( 176)$ \\
3D & $-3.66$  $( 478)$ & $-3.24$  $(   9)$ & $-3.21$  $(  29)$ & $-3.20$  $(  57)$ & $\mathbf{-3.18}$  $( 180)$ \\
4D & $-4.52$  $(1003)$ & $-4.25$  $(   9)$ & $-4.28$  $(  30)$ & $-4.20$  $(  64)$ & $\mathbf{-4.16}$  $( 183)$ \\
5D & $-5.95$  $(2971)$ & $-5.60$  $(   9)$ & $-5.55$  $(  32)$ & $-5.53$  $(  68)$ & $\mathbf{-5.50}$  $( 202)$ \\
10D & $ -$  $(-)$ & $-9.01$  $(   9)$ & $-8.93$  $(  45)$ & $-7.96$  $(  75)$ & $\mathbf{-7.84}$  $( 252)$ \\
20D & $ -$  $(-)$ & $-20.04$  $(  20)$ & $-18.10$  $(  37)$ & $-17.27$  $(  83)$ & $\mathbf{-16.11}$  $( 359)$ \\
\hline
\end{tabular}}
\end{center}
\end{table*}

A major advantage of BMDC and BMDC-T with the RAPF method for Bayesian inference is scalability.
Table \ref{tbl:timeComparison} shows prediction sensitivity of the regularized particle filter to different number of particles
and total execution times in minutes in parentheses for the BMDC-T model on the daily FX dataset with 780 observations, ordered by data dimension.
BEKK-T predictions and execution times are also provided for comparison, 
except for datasets of ten or higher dimensions where BEKK-T failed to finish in a reasonable amount of time.
BEKK-T was implemented using numerical optimization routines provided by Kevin Sheppard
\footnote{\url{http:///www.kevinsheppard.com/wiki/UCSD\_GARCH/}}.

Clearly particle filtering is much faster than the numerical optimization of the log-likelihood function used by BEKK.
In fact, particle filtering can be used for intraday trading and hedging since each sequential update
is on the order of $68*60/780\approx5$ seconds for five dimensional time series with $N = 9000$ particles.
The computational cost of the particle filter is $O(N d^{3})$ at each step, as described in Section \ref{sec:inferenceMethods}.
The computational cost for BEKK is $O(L t d^{5})$ at each step,
where $L$ denotes the average number of numerical iterations needed per cycle and
$t$ is the length of the training data currently seen.
The power $d^5$ in this cost originates because, at each iteration of the optimization process, BEKK has to compute
the gradient of $\mathcal{O}(d^2)$ parameters, where each gradient computation requires $\mathcal{O}(d^3 t)$ operations.
This cost makes BEKK infeasible for large datasets.
In contrast, the cost of the RAPF at each step does not depend on $t$ and only scales as $\mathcal{O}(d^3)$.
This allows the RAPF to analyze long high-dimensional time series.

Table \ref{tbl:timeComparison} also shows the sensitivity of the predictions of BMDC-T to the number of particles $N$.
Increasing $N$ improves performance, but has relatively small effect for low dimensional datasets.
For higher dimensions, that is, $d \geq 10$, improvements are more substantial, but do not scale linearly with $N$.
Since run times are linear in $N$, but improvements in predictive performance are not,
we can choose $N$ such that inference and prediction are done in a desired amount of time for high dimensional datasets.

\begin{table}[h]
\caption{Avg. Predictive Likelihood on Daily Equity Data}\label{tbl:BMDCDailyEquity}
\centering
\resizebox{0.475 \textwidth }{!}{
\begin{tabular}{@{\ica}l@{\ica}c@{\ica}c@{\ica}c@{\ica}c@{\ica}c@{\ica}}
Dataset &  BEKK   &  BEKK-T   &  BMDC   &  BMDC-T   \\ 
\hline
A & $-1.24$ & $\mathbf{-1.16}$ & $-1.19$ & $-1.18$ \\
AA & $-1.25$ & $\mathbf{-1.23}$ & $\mathbf{-1.24}$ & $\mathbf{-1.23}$ \\
AAPL & $-1.36$ & $-1.24$ & $\mathbf{-1.23}$ & $\mathbf{-1.23}$ \\
ABC & $-1.31$ & $\mathbf{-1.23}$ & $-1.26$ & $-1.25$ \\
ABT & $-1.30$ & $\mathbf{-1.26}$ & $-1.28$ & $-1.27$ \\
AFL & $-1.08$ & $\mathbf{-0.98}$ & $-1.01$ & $-1.01$ \\
AGN & $-1.31$ & $\mathbf{-1.26}$ & $-1.28$ & $-1.27$ \\
AIG & $-0.75$ & $\mathbf{-0.65}$ & $-0.68$ & $-0.68$ \\
AIV & $-3.00$ & $\mathbf{-0.91}$ & $-0.92$ & $\mathbf{-0.92}$ \\
AKAM & $-1.27$ & $\mathbf{-1.15}$ & $-1.20$ & $-1.20$ \\
ACE,ADSK & $-2.58$ & $\mathbf{-2.38}$ & $-2.44$ & $-2.40$ \\
ADBE,AEE & $-2.51$ & $\mathbf{-2.31}$ & $-2.37$ & $-2.36$ \\
ADI,AEP & $-2.38$ & $\mathbf{-2.29}$ & $-2.31$ & $-2.31$ \\
ADM,AES & $-2.55$ & $\mathbf{-2.21}$ & $-2.28$ & $-2.25$ \\
ADP,AET & $-2.65$ & $\mathbf{-2.37}$ & $-2.44$ & $\mathbf{-2.38}$ \\
AKS,AMGN & $-2.56$ & $\mathbf{-2.44}$ & $-2.51$ & $-2.48$ \\
ALL,AMT & $-2.11$ & $\mathbf{-1.98}$ & $-2.01$ & $-2.01$ \\
ALTR,AMZN & $-2.47$ & $\mathbf{-2.26}$ & $-2.36$ & $-2.32$ \\
AMAT,AN & $-2.37$ & $\mathbf{-2.31}$ & $-2.34$ & $\mathbf{-2.32}$ \\
AMD,ANF & $-2.61$ & $\mathbf{-2.45}$ & $-2.53$ & $-2.51$ \\
ADSK,AFL,AKS & $-5.54$ & $-3.98$ & $-3.53$ & $\mathbf{-3.43}$ \\
AEE,AGN,ALL & $-4.50$ & $-3.85$ & $-3.42$ & $\mathbf{-3.34}$ \\
AEP,AIG,ALTR & $-3.57$ & $-3.26$ & $-2.97$ & $\mathbf{-2.94}$ \\
AES,AIV,AMAT & $-4.56$ & $-3.25$ & $-3.06$ & $\mathbf{-3.03}$ \\
AET,AKAM,AMD & $-3.93$ & $-3.82$ & $-3.72$ & $\mathbf{-3.57}$ \\
AMGN,AON,APOL & $-4.06$ & $-3.60$ & $-3.71$ & $\mathbf{-3.43}$ \\
AMT,APA,ARG & $-6.07$ & $-3.49$ & $-3.35$ & $\mathbf{-3.29}$ \\
AMZN,APC,ATI & $-4.51$ & $-3.91$ & $-3.66$ & $\mathbf{-3.60}$ \\
AN,APD,AVB & $-3.42$ & $-3.41$ & $-3.25$ & $\mathbf{-3.22}$ \\
ANF,APH,AVP & $-6.04$ & $-3.70$ & $-3.68$ & $\mathbf{-3.54}$ \\
\hline
\end{tabular}}
\end{table}

\begin{table}[h]
\caption{Cumulative predictive log-likelihood for BEKK-Full, BEKK, GWP and BMDC.}\label{tbl:GWPComparison}
\begin{center}
\resizebox{0.375 \textwidth }{!}{
\begin{tabular}{@{\ica}l@{\ica}c@{\ica}c@{\ica}c@{\ica}c@{\ica}}
Dataset & BEKK-Full &  BEKK & GWP  &  BMDC  \\ 
\hline
FX (3D) & $ 2025  $  & $ 2050 $ & $2020$ & $ \mathbf{ 2130}$  \\
Equity (5D) & $ 2785 $  & $ 2800 $ & $ 2930 $ & $ \mathbf{ 3090}$  \\
\hline
\end{tabular}}
\end{center}
\end{table}

\section{Experiments for BMDC vs. GWP}\label{sec:resultsGWP}

We performed another series of experiments comparing BMDC, 
BEKK with full parameter matrices (BEKK-Full) and the diagonal version of BEKK with
the generalized Wishart process (GWP) proposed by \citet{Wilson11}.
For these experiments, we used the two financial datasets analyzed previously by these authors.
The first one corresponds to the daily returns of three currencies with respect to the US dollar:
the Canadian dollar, the Euro and the British pound. This three-dimensional time series contains 400 observations from 15/7/2008 to 15/2/2010.
This datasets is refered to as FX. The second dataset was generated from the daily returns on five equity indexes:
NASDAQ, FTSE, TSE, NIKKEI, and the Dow Jones Composite over the period from 15/2/1990 to 15/2/2010. 
A sequence of time-varying empirical covariance matrices $\tilde{\bm \Sigma_t}$ was obtained from the returns of these indexes and
then used to produce a return series by sampling from $\mathcal{N}(\mathbf{0},\tilde{\bm \Sigma_t})$ at each time step for a total of 400 steps.
This dataset is refered to as EQUITY.
In this case, the time series were not standardized to have zero mean and unit standard deviation to be consistent with \cite{Wilson11}.

We followed the same experimental protocol as in the previous section.
During the experiments, each method receives an initial time-series of length 200.
We then make predictions one step forward for 200 iterations.
In these experiments, the predictions for BMDC were generated in the same way as in \cite{Wilson11},
that is, instead of averaging $p(\mathbf{x}_t|\theta_t,\bm \Sigma_{t-1},\mathbf{x}_{t-1})$
over the available particles, we just evaluate $p(\mathbf{x}_t|\theta_t,\bm \Sigma_{t-1},\mathbf{x}_{t-1})$ on
the posterior mean estimate of $\theta_t$ and $\bm \Sigma_{t-1}$, which is approximated by the empirical mean computed across all the particles.
The predictions for GWP were done similarly, by evaluating a Gaussian likelihood on the posterior mean of $\bm \Sigma_t$, as approximated
by averaging over the samples produced by a Markov chain Monte Carlo method.

In this section we do not evaluate the performance of BMDC-T.
The reason for this is that GWP assumes a Gaussian likelihood for the data and the Student's $t$ likelihood used by BMDC-T would give an
advantage to this method with respect to GWP.

\subsection{Results for BMDC vs. GWP}

Table \ref{tbl:GWPComparison} shows the cumulative predictive log-likelihood obtained by each method on the FX and EQUITY datasets.
The method with the best predictive performance is highlighted in bold. 
In both datasets, BMDC is the best performer.
BEKK-Full underperforms diagonal BEKK, which was used as a benchmark throughout this paper.
This is likely due to worse overfitting problems in BEKK-Full, which is more highly parameterized.
GWP and the different BEKK methods have mixed performance.
GWP outperforms BEKK and BEKK-Full on the EQUITY dataset, which was generated from an empirical time-varying estimate of the return covariances.
By contrast, BEKK outperforms GWP on the real-world FX dataset.
Note the positive predictive log-likelihoods result from keeping the same experimental protocol as the GWP paper, where the return data were not rescaled to have zero mean and unit standard deviation.

\section{Conclusion} \label{sec:summary}

We have introduced a novel Bayesian Multivariate Dynamic Covariance model (BMDC) with time-varying parameters that follow a diffusion process.
The proposed model can adapt its parameters to changing dynamics in financial markets,
which results in significant improvements in prediction performance over standard econometric models
such as BEKK and other more recent methods such as the generalized Wishart process.
In addition to this, we have presented an inference method based on particle filtering
that yields substantial savings in computation time, enabling scalable inference to high-dimensional and high-frequency datasets.

\subsubsection*{Acknolewdgements}

JMHL is supported by Infosys Labs, Infosys Limited.

\bibliography{example_paper}

\bibliographystyle{icml2013}

\end{document}